\documentclass[aps,prl,reprint,groupedaddress,showpacs]{revtex4-1}

\usepackage{amssymb,amsmath}
\usepackage{graphicx}
\usepackage{color}

\begin{document}


\title{Quench-induced supercurrents in an  annular Bose gas}


\author{L. Corman$^{1}$}
\author{L. Chomaz$^{1,4}$}
\author{T. Bienaim\'e$^{1}$}
\author{R. Desbuquois$^{2}$}
\author{C. Weitenberg$^{3}$}
\author{S. Nascimb\`ene$^{1}$}
\author{J. Dalibard$^{1,4}$}
\author{J. Beugnon$^{1}$}

\email[]{beugnon@lkb.ens.fr}
\affiliation{$^1$Laboratoire Kastler Brossel, CNRS, UPMC, ENS, Coll\`ege de France, 24 rue Lhomond,  75231 Paris Cedex 05, France}
\affiliation{$^2$Institut f\"ur Quantenelektronics, ETH Zurich, 8093 Zurich, Switzerland}
\affiliation{$^3$Institut f\"ur Laserphysik, Universit\"at Hamburg, Luruper Chaussee 149, D-22761 Hamburg, Germany}
\affiliation{ $^4$Coll\`ege de France, 11 Place Marcelin Berthelot, 75005, France}


\date{\today}
\begin{abstract}
We create supercurrents in annular two-dimensional Bose gases through a temperature quench of the normal-to-superfluid phase transition. We detect the magnitude and the direction of these supercurrents by measuring spiral patterns resulting from the interference of the cloud with a central reference disk. These measurements demonstrate the stochastic nature of the supercurrents. We further measure their distribution for different quench times and compare it with predictions based on the Kibble-Zurek mechanism.
\end{abstract}

\maketitle

Fluids in annular geometry are ideally suited to investigate
the role of topological numbers in quantum mechanics. The phase winding
of the macroscopic wavefunction around the annulus must be a multiple
of $2\pi$, ensuring the quantization of the circulation
of the fluid velocity. The resulting supercurrents have been observed
in superfluid systems such as superconductors \cite{Silver1967},
liquid helium \cite{Bendt1962} and atomic gases
\cite{Ryu2007,Moulder2012}. Studying these currents is crucial for
the understanding of quantum fluids, as well as for realizing sensitive detectors like
magnetometers \cite{Clarke1977} and rotation sensors \cite{Packard1992}.

Supercurrents in annular atomic Bose-Einstein condensates (BECs) are usually created
in a deterministic way by using laser beams to impart angular momentum on the atoms \cite{Ryu2007,Moulder2012,Beattie2013}  or  by rotating a weak link along the annulus \cite{Wright2013}. Supercurrents can also have a stochastic origin. They may result from thermal fluctuations or appear as topological defects following a rapid quench of the system. The latter mechanism was put forward by Kibble and Zurek (KZ), who
studied the phase patterns that emerge in a fluid, when it undergoes
a fast crossing of a phase transition point \cite{Zurek1985,Kibble1976}.

The KZ mechanism has been studied in several types of systems such
as liquid crystals \cite{Chuang1991}, helium \cite{Ruutu1996,Bauerle1996},
ion chains \cite{Ulm2013,Pyka2013}, superconducting loops \cite{Monaco2009} and BECs \cite{Weiler2008,Lamporesi2013,Braun2014}.
For a superfluid confined in a ring geometry, which is the configuration
originally considered by Zurek \cite{Zurek1985}, the frozen phase of the wavefunction may lead to a supercurrent of charge $q$, i.e. a $2 \pi q$ phase winding along the ring. 
In this Letter, we study a setup realizing this \emph{gedanken} experiment using a quasi two-dimensional (2D) Bose gas trapped in an annular geometry. For each realization of the experiment, we use matter-wave interference between this annulus and
a central disk acting as a phase reference, to measure the charge
as well as the direction of the random supercurrent \footnote{A similar method has recently been developed to investigate the supercurrent generated by a rotating weak link \cite{Eckel2014}}.

Our experiments are performed with a Bose gas of $^{87}$Rb atoms. 
Along the vertical ($z$) direction the gas is confined using a harmonic potential with frequency $\omega_{z}/2\pi=370$\,Hz (figure \ref{fig1}a) \footnote{\label{ulm}For details, see Supplemental Material.}. In the horizontal ($xy$) plane, the atoms are trapped in the dark
regions of a ``box-potential'' beam, engineered using an intensity
mask located in a plane optically conjugated to the atom cloud \footnote{see \cite{Gaunt2013} for a 3D version of a similar setup}. We
use a target-like mask, consisting of a disk of radius $R_{0}=4.5$\,$\mu$m
surrounded by a ring of inner (resp. outer) radius of $R_{\rm in}=9$\,$\mu$m
(resp. $R_{\rm out}=15$\,$\mu$m) (figure \ref{fig1}b). 

\begin{figure}[htb!]
\begin{centering}
\includegraphics[width=8.6cm]{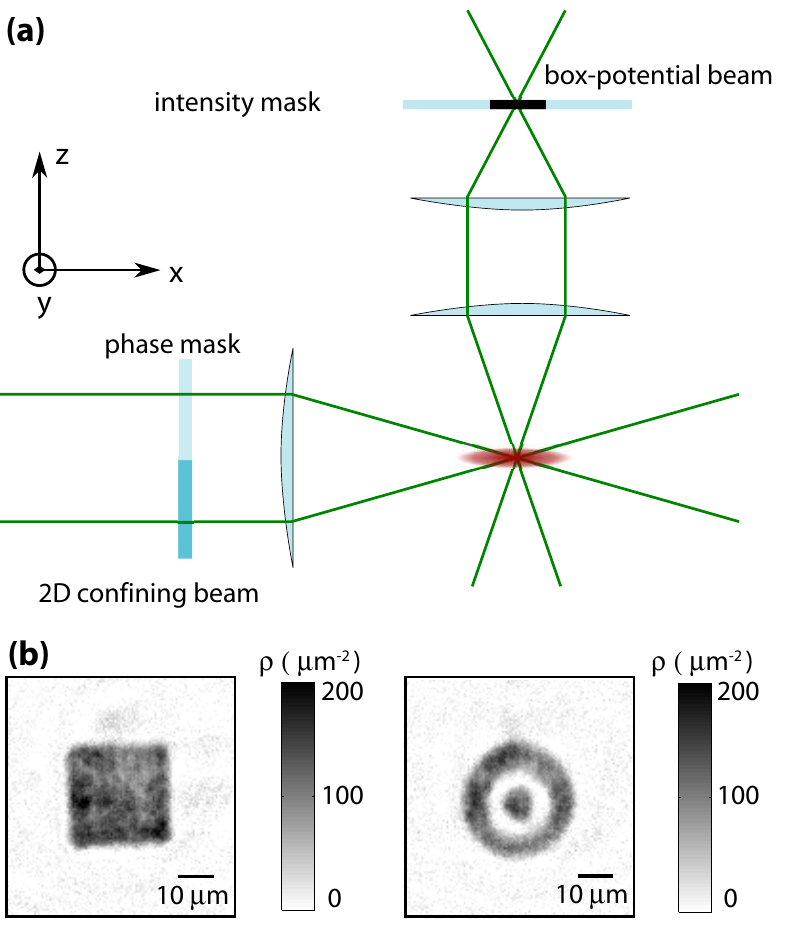} 
\end{centering}
\caption{Production of box-like potentials using an intensity mask. (a) Along the vertical direction, atoms are confined by a laser beam with an intensity node in the plane $z = 0$ which is shaped using a phase plate ($\pi$ phase shift between the upper and lower halves of the phase plate). In plane, atoms are trapped in box-like potentials created by imaging an intensity mask onto the atomic plane. The box-like potentials are created by imaging an intensity mask onto the atomic plane. (b) In-situ images of uniform gases in the square and target potentials.\label{fig1}}
\end{figure}

\begin{figure}[hbt!]
\begin{centering}
\includegraphics[width=6cm]{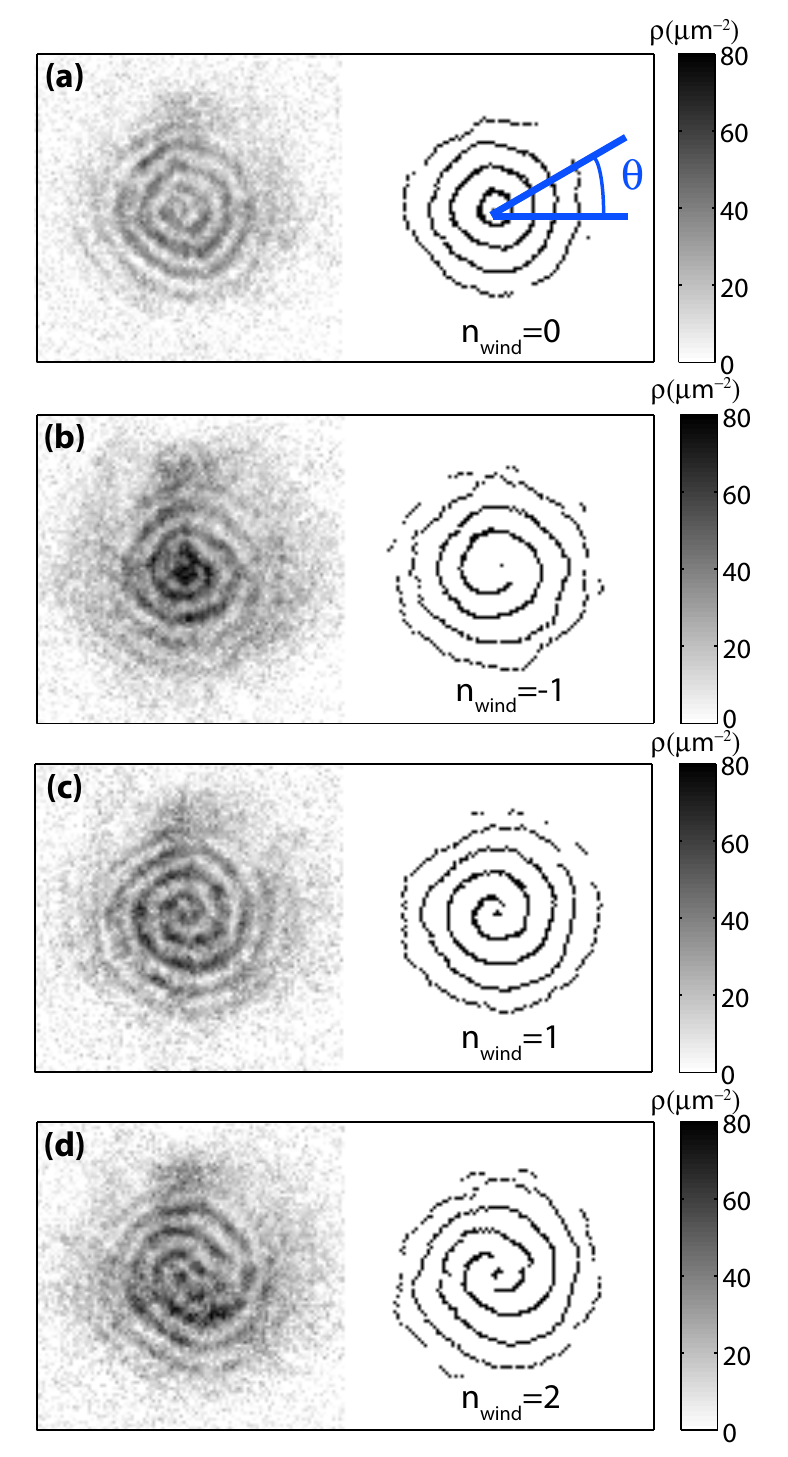} 
\par\end{centering}
\caption{Experimental interference patterns. Examples of interference patterns after expansion in the 2D plane, along with constrast-amplified pictures. (a) without phase winding, (b) with phase winding $-2\pi$, (c) with phase winding $+2\pi$, (d) with phase winding $+4\pi$.}\label{fig2}
\end{figure}

The typical time sequence for preparing the gas starts by loading a gas with a 3D phase-space density $\approx$ 2.4 slightly below the condensation threshold \footnote{The estimated total atom number 76000 and the temperature is 210 nK. With these parameters, we never observe any interference fringes such as those of fig \ref{fig2}.} with the box-potential
beam at its maximal power. Then we lower linearly this power by a factor $\sim50$
in a time $t_{{\rm evap}}$ to evaporatively cool the atomic cloud
and cross the superfluid transition \cite{Desbuquois2012}. Last
we keep the atoms at a constant box potential depth during a time $t_{{\rm hold}}$.
The final temperature is $\sim10$\,nK with similar surface densities in the ring and the disk: $\rho\sim$\,80\,$\mu$m$^{-2}$. The typical interaction energy per atom is $E_{{\rm int}}/k_{{\rm B}}\approx 8$\,nK, and the gas is marginally  quasi-2D with $k_{{\rm B}}T,\: E_{{\rm int}}\sim\hbar\omega_{z}$. These parameters correspond to a large 2D phase-space density, ${\cal D}=\rho\lambda^{2}\geq 100$, so that the gas is deeply in the superfluid regime at the end of the evaporation ramp ($\lambda=\left[2\pi\hbar^{2}/(mk_{{\rm B}}T)\right]^{1/2}$  is the thermal wavelength and $m$ the mass
of the $^{87}$Rb atom).

We use matter-wave interference to probe the relative phase distribution
between the cloud in the central disk and the one in the ring. We
abruptly switch off the box-potential while keeping the confinement
along the $z$ direction. The clouds experience a hydrodynamical expansion
during which the initial interaction energy is converted into kinetic
energy. After $7\,$ms of expansion, we record the interference pattern
by imaging the atomic gas along the vertical direction.
Typical interference patterns are shown in figure \ref{fig2}.
Most of them consist in concentric rings, as expected for a quasi-uniform
phase distribution in the disk and the annulus. However we also observe
a significant fraction of spiral patterns, revealing the presence of a phase winding in the wavefunction of one of the two clouds. 

\begin{figure}[hbt!]
\begin{centering}
\includegraphics[width=8.6cm]{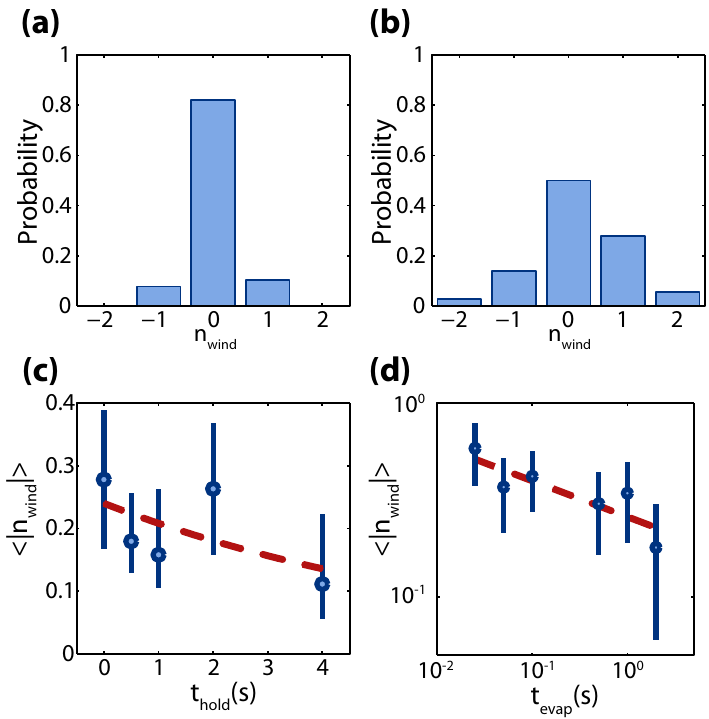} 
\par\end{centering}
\caption{Study of the winding number. (a) and (b) Histograms showing the statistical appearance of winding number $n_{{\rm wind}}$ for $t_{{\rm hold}}=0.5$\,s. (a) We show the result of 39 realizations for $t_{{\rm evap}}=2$\,s. We get $\langle n_{{\rm wind}}\rangle=0.03\,(8)$. (b) We show the result of 36 realizations for $t_{{\rm evap}}=0.025$~s. We get $\langle n_{{\rm wind}}\rangle=0.19\,(14)$. (c) Mean absolute winding number $\langle|n_{{\rm wind}}|\rangle$ as a function of hold time ($t_{{\rm evap}}=2\,$s). The data is fitted with an exponential with a time constant of 7\,s.
(d) Mean absolute winding number $\langle|n_{{\rm wind}}|\rangle$ as a function of evaporation time
($t_{{\rm hold}}=0.5\,$s) in log-log scale. The line is a power-law fit to the data, $\langle|n_{{\rm wind}}|\rangle\propto t_{{\rm evap}}^{-\alpha}$, gives $\alpha=0.19\,(6)$. The uncertainty on $\langle n_{{\rm wind}}\rangle$ and the bars on figure 3c-d represent the statistical error determined with a bootstrapping approach described in Supplemental Material.
\label{fig3}}
\end{figure}

We developed an automatized procedure to analyze these patterns, which
reconstructs the phase $\phi(\theta)$ of the fringes along a line
of azimuthal angle $\theta$ (see Supp. Mat.). From the accumulated phase $\Delta\phi$ as the angle $\theta$
varies from 0 to $2\pi$, we associate to each pattern a winding number
$n_{{\rm wind}}=\Delta\phi/2\pi$, which is a positive, null or
negative integer. This number is recorded for many realizations of the same experimental sequence. Examples of the probability distribution of $n_{\rm wind}$ are shown on figure \ref{fig3}a and b. 
The measured histograms are compatible with a zero mean value \footnote{The observed asymetry on figure \ref{fig3}b (mean value is 1.4 times the standard deviation) is compatible with the number of realizations:  the probability to have a standard deviation equal or larger than this one is 17\%.}. For example, if we use all the data presented on figure \ref{fig3}c and \ref{fig3}d we find $\langle n_{{\rm wind}}\rangle=0.002\,(20)$. This confirms the stochastic nature of the mechanism at the origin of this phase winding.


The first question that arises is the origin of the observed phase winding,
which can be due either to a vortex in the central disk or to a quantized
persistent current in the outer ring. We can experimentally eliminate
the first possibility by noticing that when doing a 3D ballistic expansion (by switching-off both the box-potential beam and the confining beam in the $z$ direction)  we never observe any vortex signature in the small disk of radius $R_0=4.5\,\mu$m. By contrast, in larger structures such as the square represented in Fig. \ref{fig1}b , we can detect deep density holes revealing the presence of vortices \cite{Chomaz2014}. Hence we conclude that the spiral interference patterns of figure \ref{fig2} reveal the presence
of a supercurrent in the annulus, whose charge and orientation correspond
to the modulus and sign of the winding number $n_{{\rm wind}}$. 
The lifetime of this supercurrent is similar to the cloud lifetime (see Fig. \ref{fig3}c).

We now discuss the origin of the observed supercurrents,
which can be either thermal excitations or result from the quench
cooling. If these currents had a thermal origin, their probability
of occurrence would be given by the Boltzmann law $p(n_{{\rm wind}})\propto\exp\left[-E(n_{{\rm wind}})\,/\, k_{{\rm B}}T\right]$,
where the (kinetic) energy of the supercurrent is 
\begin{equation}
E(n_{{\rm wind}})=n_{{\rm wind}}^{2}\,\frac{\pi\hbar^{2}\rho}{m}\,\ln\left(R_{\rm out}/R_{\rm in}\right).\label{eq:vortex_energy}
\end{equation}
This leads to 
\begin{equation}
p(n_{{\rm wind}})\propto\left(R_{\rm in}/R_{\rm out}\right)^{n_{{\rm wind}}^{2}{\cal D}/2},\label{eq:pnwind}
\end{equation}
 which is negligible for $n_{{\rm wind}}\neq0$ for our large phase
space densities ${\mathcal{D}} \geq 100$, in clear disagreement with the typical
20-50\% of pictures showing phase winding. Note that the probability
for a vortex to appear in the central disk as a thermal excitation
is even smaller than (\ref{eq:pnwind}) because $R_{\rm in}$ and $R_{\rm out}$ should be replaced respectively by the healing length ($\lesssim 0.5\,\mu$m) and $R_{0}$.

To check that the quench cooling is indeed responsible for the formation
of these supercurrents, we study the variation of $\langle|n_{{\rm wind}}|\rangle$
for evaporation times spanning two orders of magnitude. The comparison between the results for a slow quench (figure \ref{fig3}a) and a fast quench (\ref{fig3}b) show that the latter indeed increases the probability of occurrence of a supercurrent, as expected for the KZ mechanism \cite{Zurek1985,Kibble1976}. We summarize in figure \ref{fig3}d the experimental variation of $\langle|n_{{\rm wind}}|\rangle$ with  $t_{{\rm evap}}$, and find that   it increases from 0.2 ($t_{\rm evap}=2$\,s) to 0.6 ($t_{\rm evap}=0.025$\,s). A power-law fit to the data, inspired by the prediction for the KZ mechanism, leads to $\langle|n_{{\rm wind}}|\rangle\propto t_{{\rm evap}}^{-\alpha}$ with $\alpha=0.19\,(6)$. 

To interpret our results we have developed a simple one-dimensional (1D) model following the KZ scenario presented in \cite{Zurek1985,Das2012}. We consider a 1D ring of perimeter $L$ and we assume that, when the normal-to-superfluid transition is crossed, $N$ domains of uniform phase $\phi_j$, $j=1,\ldots,N$ are created. Each run of the experiment is modeled by a set $\{ \phi_j\}$, where the phases $\phi_j$ are independent random variables drawn in $(-\pi,\pi]$ (with $\phi_1=0$ by convention). For each set of  $\{\phi_j\}$ we calculate the total phase variation along the ring $\Phi=\sum_j \phi_j$ and define $n_{\rm wind}$ as the nearest integer to $\Phi/2\pi$. We then average over many draws of the set $\{\phi_j\}$. Our experimental range  $0.2 \leq \langle|n_{\rm wind}|\rangle\leq 0.6$  is obtained for $3\leq N\leq 10$, corresponding to the approximate power-law scaling (see Supp. Mat.)
\begin{equation}
\langle|n_{\rm wind}|\rangle \propto N^{0.8} .
\label{eq:nwind_exp}
\end{equation}
Then we use the general prediction for the KZ mechanism to relate the typical length $\hat \xi=L/N$ of a domain to the quench time $t_{\rm evap}$ (see e.g. \cite{Das2012})
\begin{equation}
\hat{\xi}\propto t_{\rm evap}^{\nu/(1+\nu z)},
\label{eq:power_law}
\end{equation}
where $\nu$  and $z$ define the universality class of the transition: $\nu$  is the correlation length critical exponent and $z$ the dynamic critical exponent. Using $z=2$ and $\nu=1/2$ relevant for a mean-field description of a 1D ring-shaped system \cite{Das2012}, we get 
\begin{equation}
\hat{\xi}=\frac{L}{N}\propto t_{\rm evap}^{1/4}.
\label{eq:xi}
\end{equation}
Combining (\ref{eq:nwind_exp}) and (\ref{eq:xi}), we predict with this simple model
\begin{equation}
\langle|n_{\rm wind}|\rangle\propto t_{\rm evap}^{-1/4\times0.8}\approx t_{\rm evap}^{-0.2},
\end{equation}
which is in agreement with the experimental result $\alpha=0.19\,(6)$. 

There are two main assumptions that could limit the validity of this model. First, our system is not uni-dimensional in terms of relevant single particle eigenstates. However,  we find for our parameters that $\hat \xi$ is in the range $7$--25\,$\mu$m \footnote{ An estimate of $\hat{\xi}$ for our geometry is $\pi\, (R_{\rm in}+R_{\rm out})/N$}\footnote{We note that $\hat{\xi}$ is then larger than the size $R_0$ of the central disk. This confirms the fact that we do not expect the presence of vortices in this disk}; this is always larger than the width of our annulus and justifies the use of a 1D model for describing the phase coherence properties of the gas. Second, this model does not take into account beyond mean-field effects, related to either the finite size of the system or the crossover between standard BEC and the Berezinskii--Kosterlitz--Thouless mechanism. This could change the value of the critical exponents and even lead to deviations with respect to the power-law scaling of (\ref{eq:power_law}) \cite{Jelic2011}.

We now discuss the possible extension of this work to a more thorough test of the KZ mechanism. Power-law scaling is challenging to test in our situation because of the low value of the exponent ($\approx 0.2$) even if we span two orders of magnitude for $t_{\rm evap}$. The extreme values of this range are experimentally limited: (i) The evaporation time $t_{\rm evap}$ should be chosen long enough so that at any given time a local thermal equilibrium is achieved in the cloud (see Supp. Mat.). (ii) The largest evaporation time is set by the cloud lifetime. These two limits cannot be significantly modified, which fixes the relative range of variation of the number of domains $N$. It could also be interesting to study situations with absolute larger $N=L/\hat{\xi}$. For a given density the local equilibrium requirement limits the lower value of $\hat{\xi}$ and one can only increase the length of the ring $L$. Within current experimental techniques, it should be possible to load one order of magnitude more atoms, leading for a given transverse size to an increase of $N$ by the same factor.

\begin{figure}[t]
\begin{centering}
\includegraphics[width=8.6cm]{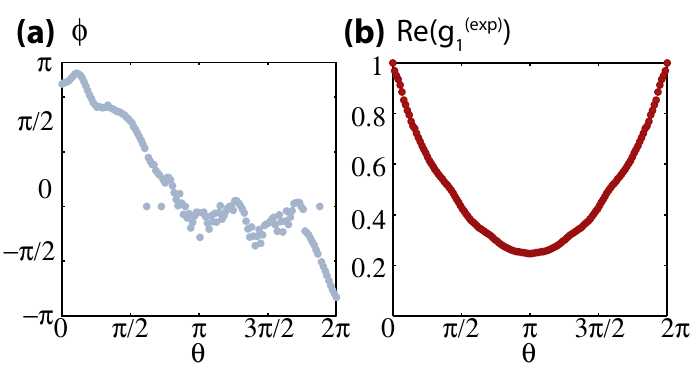} 
\par\end{centering}
\caption{Analysis of the phase profiles. (a) Typical phase distribution reconstructed
from the phase profile $\phi(\theta)$ of the interference pattern of figure \ref{fig2}b showing a winding number of -1. (b) Real value of the angular correlation function reconstructed from the
phase of the interference patterns with 18 realizations of $t_{{\rm evap}}=2\,$s
and $t_{{\rm hold}}=4\,$s. When $n_{\rm wind}\neq0$ the linear phase winding is substracted before computing $g_{1}$. \label{fig4}}
\end{figure}

In the last part of this Letter, we show that one can extract information from the interference patterns, which goes beyond  the determination of the topological number $n_{{\rm wind}}$.  In particular the ripples of the fringes
are related to the phase distribution of the fluids in the central
disk and the ring, which is characterized by the one-body correlation
function $g_{1}$.  
This function plays a specially important role for low-dimensional
systems, since it indicates how long-range order is destroyed by thermal
phonons.  To give an estimate of $g_{1}$, we study the angular dependance of the phase of the fringes $\phi(\theta)$ as shown on figure \ref{fig4}a. In particular we consider the periodic function $\delta\phi(\theta)=\phi(\theta)-n_{{\rm wind}}\,\theta$, which describes the deviation of the reconstructed phase from a perfect linear winding.
We construct the angular correlation function: 
\begin{equation}
g_{1}^{{\rm (exp)}}(\theta)=\langle{\rm e}^{{\rm i}\left[\delta\phi(\theta')-\delta\phi(\theta'+\theta)\right]}\rangle_{\theta',\,{\rm realizations}},\label{eq:one-body-density-matrix}
\end{equation}
 where the average is taken over all images irrespective of the value
of $n_{{\rm wind}}$, and which is expected to be real in the limit
of a large number of realizations. A typical example for ${\cal R}{\rm e}[g_{1}^{{\rm (exp)}}]$
is given in figure \ref{fig4}b, where the minimum for $\theta=\pi$
gives an indication of the phase coherence between diametrically opposite
points. This measured  angular correlation function $g_{1}^{{\rm (exp)}}(\theta)$ can be used to reconstruct the first-order correlation function of the gas in the annulus (see Supp. Mat.). This correlation function could allow one to extract the evolution of the phonon distribution during the thermalization of the fluid.

In summary, we have created supercurrents in annular Bose gases by a temperature quench. The measured distribution of direction and magnitude of these supercurrents are compatible with the KZ mechanism's predictions. This work could be extended to more refined tests of the KZ mechanism by testing the power-law scaling with the size of the annulus and correlate the number of topological defects with the condensed fraction of the system \cite{Das2012}. \\

\appendix

\section*{Appendix A: Methods}
\begin{figure}[ht]
\begin{centering}
\includegraphics[width=8.6cm]{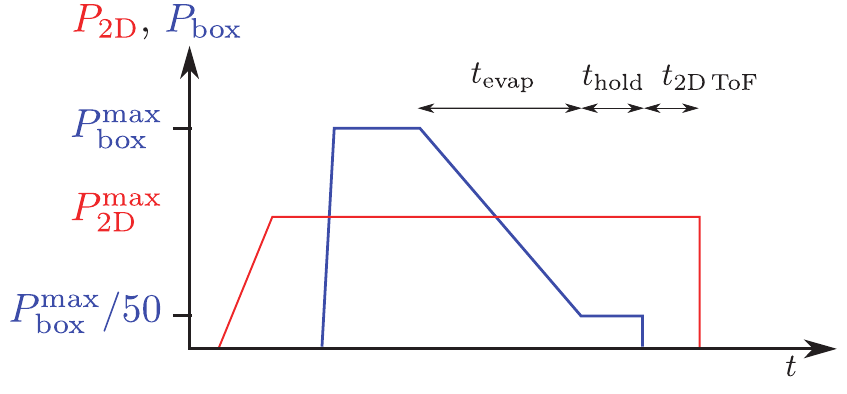} 
\par\end{centering}
\caption{Experimental sequence: After ramping up the power of the 2D-confining beam, we switch on the box-potential beam and then divide the power of the beam by 50 in a time $t_{\rm evap}$ and keep it constant for a time  $t_{\rm hold}$ after which we release the atoms from the box-potential to reveal the interference patterns.}\label{figS3}
\end{figure}

\paragraph*{\textbf{Preparation of the cloud}}
We prepare a Bose-Einstein condensate of $^{87}$Rb in a combined
magnetic and optical potential as described in \cite{Lin2009a}. After
radio-frequency evaporation in a quadrupole trap, the cloud is transferred
into the dipole trap. The residual quadrupole field provides both
confinement along the propagation axis of the dipole beam and partial
gravity compensation for atoms in the $|F=2,m=2\rangle$ electronic ground state ($g_{{\rm eff}}=0.1\, g$ where $g$ is the acceleration
of gravity). We evaporate further by lowering the power of the dipole
trap to reach a condensate of $10^{5}$ atoms. The following time sequence is described on figure \ref{figS3}. First, the 2D-confining beam is ramped up in 1~s. It consists in a
beam at 532~nm with power 300~mW shone on a phase plate that imprints
a $\pi$ phase shift to its upper half. In the far field, the diffracted field is approximately zero along the $y$ axis, so that it creates a repulsive dipole potential on the atoms with a minimum in the $z=0$ plane around which the atoms are confined.
The waists on the atoms (without phase plate) are $w_{z}=11$~$\mu$m
and $w_{y}=50$~$\mu$m. We can load the whole cloud into this trap.
The frequency of the harmonic potential along the vertical direction $z$ is $\omega_{z}/2\pi=370$~Hz.
The anticonfinement due to the 2D beam in the $y$ direction (orthogonal
to the propagation axis) is equivalent to an inverted harmonic potential
with frequency $\omega_{{\rm anti}}/2\pi= 4.2$~Hz. We then ramp up the box-potential beam in 0.1~s to its maximal power
(500~mW). The cloud is held for 150~ms, allowing for the atoms outside
of the box-potential to fall. The power of the box-potential beam
is then lowered within a time $t_{\rm evap}$ to its final value, corresponding to a barrier height of 45~nK, and kept constant for a time $t_{{\rm hold}}$. At this stage we obtain a uniform 2D gas (the production of 3D uniform gases in similar ``optical box potentials" has recently been reported in \cite{Gaunt2013}). 
After the hold time, the atoms are released from the box-potential
trap and gravity is fully compensated. The atoms are let to expand
for a typical time of 7~ms. The cloud is then imaged using high-intensity
imaging \cite{Reinaudi2007} to avoid multiple scattering effects
in the imaging of dense 2D gases \cite{Chomaz2012}. The optical resolution
of our imaging system is on the order of 1~$\mu$m.

\paragraph*{\textbf{Uniformity of the gas}}
The resolution of the optical system projecting
the image of the mask onto the atoms is 3~$\mu$m. The steepness
of the resulting potential barrier, defined as the length scale over
which the box potential varies from 5\% to 25\% of its maximal value,
is 1~$\mu$m. The root-mean-square variation of the beam intensity in the central region is 3\% of the total height of the barrier, which allows us to obtain uniform gases at
low power of the box-potential beam.

\paragraph*{\textbf{Determination of the temperature and density of the cloud}}
From the temperature fits of thermal clouds in disk-shaped box-potentials,
we estimate that the temperature of the clouds corresponds to a fraction
$\eta$ of the height of the barrier. Typically,
the measured value of $\eta$ is 4. Thus we estimate a temperature of $\approx $10~nK
for the clouds.
The atom density is determined by averaging high-intensity \cite{Reinaudi2007} in-situ pictures over the size of the cloud.

\paragraph*{\textbf{Determination of the interaction energy of the cloud}}
For the ground state of the many-body system in the mean-field approximation, the total interaction energy is $E_{\rm int}=\frac{1}{2}\int d^3r\, U_{\rm int} \,n^2( \mathbf{r})$, where $n( \mathbf{r})$ is the 3D density and $U_{\rm int}=\frac{4\pi a \hbar^2}{m}$, with $a$ the $s$-wave scattering length. Assuming a uniform trapping in the horizontal plane and a harmonic confinement along the vertical direction with $\omega_z/2\pi=370$\,Hz,  we calculate the ground state of the system by solving numerically the 3D Gross-Pitaevski equation for a disk-shaped trap of $R=12 \mu$m with $N_{\rm at}=36\,000$ atoms (corresponding to the same surface density as in the experiments presented here: $\rho=80 \mu$m$^{-2}$). We find $E_{\rm int}/N_{\rm at}\approx k_B \times 8$ nK. We also checked that the residual anticonfinement along the $y$ direction has little influence on the equilibrium distribution.

\paragraph{\textbf{Statistical analysis}}

Each data point in figure 3 of the article 
is the average of 15 to 50 realizations. Error bars
for the mean absolute winding number are obtained using a bootstrapping
approach. From the initial set of data, $10\,000$ draws with replacement
of datasets with the same length as the initial sample are made. For
each draw, the mean absolute winding number is calculated. Then, using
the bias corrected and accelerated percentile method \cite{Efron1987},
the one-standard deviation confidence interval is calculated.

\paragraph{\textbf{Evaporation and temperature quench}}
The fastest ramp of the box-potential beam we use (25 ms) is still slow enough for the evaporation process to take place and to identify this ramping down as a quench of the temperature of the system: we calculated the typical elastic collision time when crossing the transition to be a few milliseconds \cite{Ketterle1996}.

\paragraph{\textbf{Reconstruction of the phase profile}}

For each picture, the center is determined manually. The shot-to-shot variation of this center is small ($\approx 0.5\, \mu$m) and comparable to the independently measured position stability
of the initial cloud. We checked that such an offset on the center does not lead to large modification of the results. Then we proceed
in two steps to reconstruct the phase profile, contrast amplification
and fit. To amplify the contrast, the pictures are first convoluted by a $2\times2$
matrix with constant coefficients. This filters out high frequency
noise but does not blur the interference pattern. Then radial cuts
with angle $\theta\in\{0,2\pi/n,\cdots,2\pi(1-1/n)\}$ are performed
(typically $n=150$), and the positions of local maxima are recorded,
giving the contrast amplified picture.

To retrieve the phase, we perform a convolution of the contrast amplified picture with a gaussian of width 3 pixels and we fit radial cuts of the convoluted, contrast-amplified
pictures with the following function 
\[
f(r,k,\phi,A,c)=A\sin\left(kr+\phi\right)+c
\]
 for points with distance to the center $r\in\left[r_{{\rm min}},r_{{\rm max}}\right]$.
First, the parameter $k$ is left as a free parameter to fit the radial
cuts. Then the averaged $k_{{\rm mean}}$ over all fits is taken as
a fixed parameter and all the radial cuts are fitted again. The phase
$\phi$ is recorded as a function of the angle $\theta$ of the radial
cut.

\section*{Appendix B:  Scaling of the mean winding number versus $t_{\rm evap}$}
\begin{figure}[ht!]
\begin{centering}
\includegraphics[width=8.6cm]{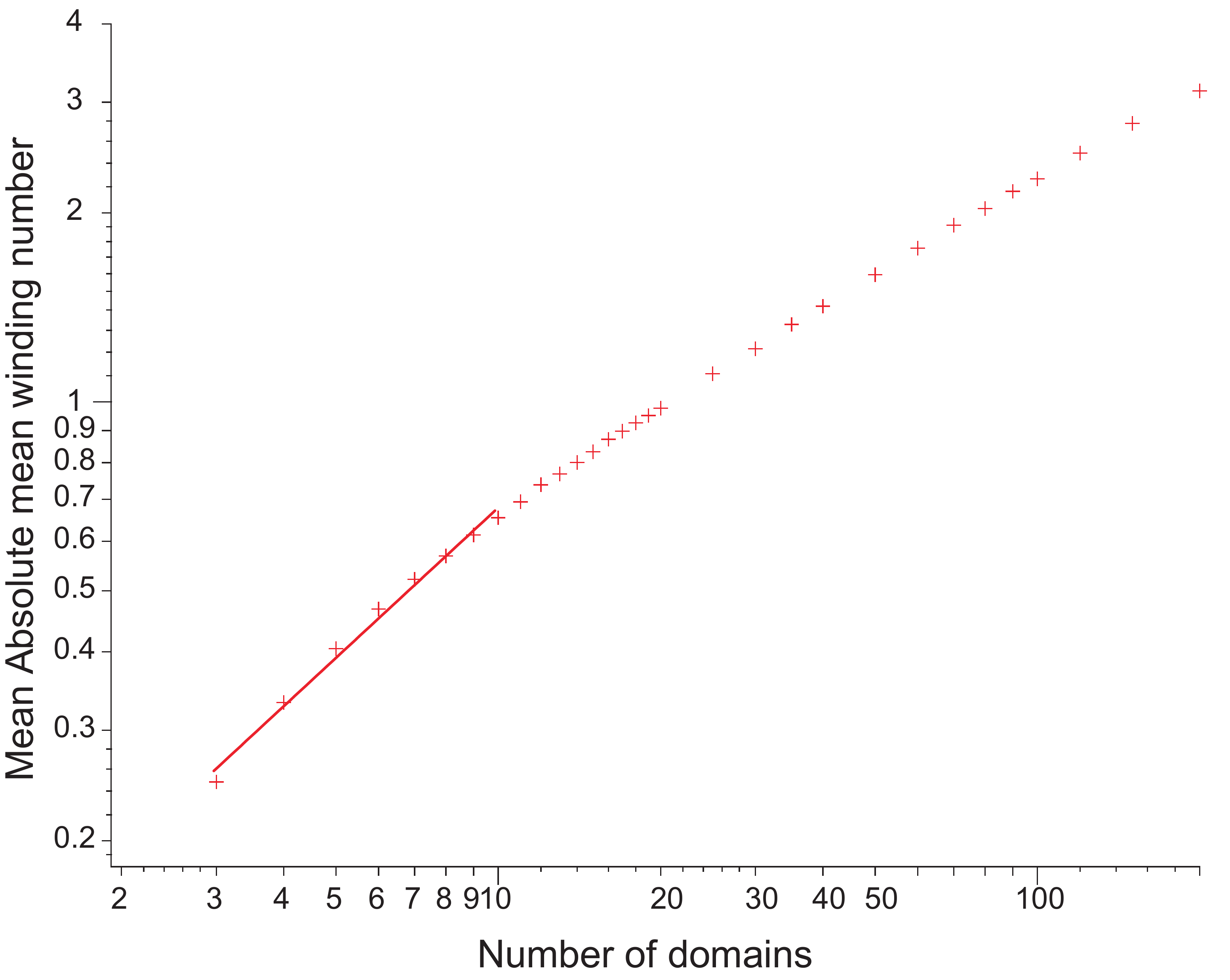} 
\par\end{centering}
\caption{ Average absolute winding number as a function of the number of phase domains in log-log scale. The points are the results of the simulation and the line is the power-law fit to the relevant points for the experiments described here.}\label{fig:S1}
\end{figure}

We report on figure \ref{fig:S1} the result of the calculation  described in the article  of the average absolute winding number $\langle|n_{\rm wind}|\rangle$ obtained as a function of $N$ the number of domains with different phases. For large values of $N$ we find that $\langle|n_{\rm wind}|\rangle$  scales like $\sqrt{N}$ as expected for a sum of a large number of independent random variables. For $3\leq N\leq 10$, we do not expect to recover an exact power-law behavior but we can still fit a power-law scaling to our data and get
\begin{equation}
\langle|n_{\rm wind}|\rangle \propto N^{0.8}.
\end{equation}

\section*{Appendix C:  First-order correlation function along the annulus.}
\begin{figure}[bht!]
\begin{centering}
\includegraphics[width=8.6cm]{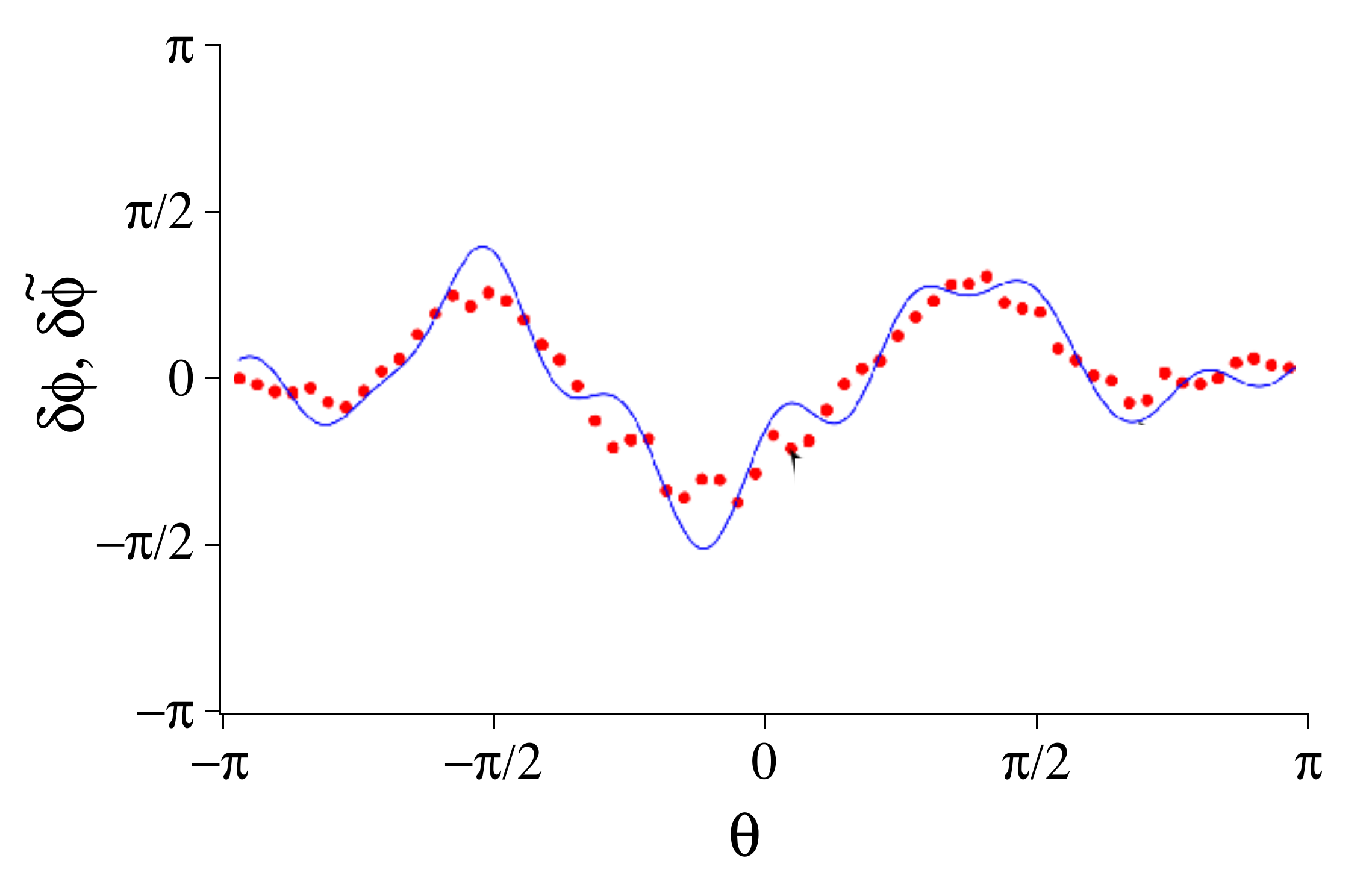} 
\par\end{centering}
\caption{Comparison between the phase fluctuations of an initial state in the annulus ($\delta \tilde{\phi}$, solid line) and the phase profile deduced with the interference method used in the experiment ($\delta \phi$, red points).}\label{fig:S2}
\end{figure}

To relate quantitatively $g_{1}^{{\rm (exp)}}(\theta)$ to the coherence
properties of  the gas in the ring, two hypotheses are needed:
(i) We suppose that the fluid in the central disk acts as a phase reference, so that the ripples of the fringes come essentially from the phase fluctuations in the ring. Indeed the small size of this disk guarantees that phonon modes are only weakly populated.
(ii) We assume that the fluctuations of the phase of the fringe pattern
directly reflect the phase of the atomic wave function along the ring.
This is validated by the following numerical analysis, in which we simulate numerically the hydrodynamical expansion and calculate the wavy interference pattern originating from a given phase distribution along the ring. We use a spatial
grid of size $36\,\mu{\rm m}\times36\,\mu{\rm m}$ with pixel size $0.52\,\mu$m.
We first compute the ground state of $N_{\rm at}=5\times10^{4}$ atoms in the
target potential using the Gross-Pitaevskii equation, evolved with
the split-step method in imaginary time (time step $10\,\mu$s). A
phase fluctuation $\delta\tilde{\phi}(\theta)$ is then added by hand to the
wave function in the ring. We then simulate the hydrodynamical expansion
by evolving the Gross-Pitaevskii equation in real time (time step $10\,\mu$s)
during $7\,$ms. The phase $\delta\phi(\theta)$ of the fringe pattern
is finally obtained using the same procedure as for experimental
pictures. A comparison between typical phase distributions $\delta\tilde{\phi}(\theta)$
and $\delta\phi(\theta)$ is given in figure  \ref{fig:S2}. Both phase profiles are similar confirming that the phase reconstructed from the interference pattern correspond to the in-situ phase of the gas

\begin{acknowledgments}
We thank J. Palomo and D. Perconte for the realization of the intensity masks and Z. Hadzibabic for useful discussions. This work is supported by IFRAF, ANR (ANR-12-BLANAGAFON),
ERC (Synergy UQUAM). L. Chomaz and L. Corman
acknowledge the support from DGA, and C. Weitenberg
acknowledges the support from the EU (PIEF-GA-2011-
299731).

LCo. and LCh. contributed equally to this work.
\end{acknowledgments}

\bibliography{vortexbibv3}

\end{document}